# A new technique for ultra-fast physical random number generation using optical chaos

Amr Elsonbaty[a,b], Salem F. Hegazy [a,c], Salah S. A. Obayya[a*]

[a]Centre for Photonics and Smart Materials, Zewail City of Science and Technology, Sheikh Zayed District,12588 Giza, Egypt.;
[b]Mathematics and Engineering Physics Department, Faculty of Engineering, Mansoura University, 35516 Mansoura, Egypt;
[c]National Institute of Laser Enhanced Sciences, Cairo University, 12613 Giza, Egypt.

## ABSTRACT

In this paper, we numerically demonstrate a new extraction scheme for generating ultra-fast physically random sequence of bits. For this purpose, we utilize a dual-channel optical chaos source with suppressed time delayed (TD) signature in both the intensity and the phase of its two channels. The proposed technique uses $M$ 1-bit analog-to-digital converters (ADCs) to compare the level of the chaotic intensity signal at time $t$ with its levels after incommensurable delay-interval $T_m$, where $m = \{1,2,...,M\}$. The binary output of each 1-bit ADC is then sampled by a positive-edge-triggered D flip-flop. The clock sequence applied to the flip-flops is relatively delayed such that the rising edge of the clock triggering the $m$ flip-flop precedes the rising edge of the clock of a subsequent $m+1$ flip-flop by a fixed period. The outputs of all flip-flops are then combined by means of a parity-check logic. Numerical simulations are carried out using values of parameters at which TD signature is suppressed for chosen values of setup parameters. The 15 statistical tests in Special Publication 800-22 from NIST are applied to the generated random bits in order to examine the randomness quality of these bits for different values of $M$. The results show that all tests are passed from $M = 1$ to $M = 39$ at sampling rate up to 34.5 GHz which indicates that the maximum generation rate of random bits is 2.691 Tb/sec using a chaotic source of single VCSEL and without employing any pre-processing techniques.

## 1. INTRODUCTION

The random number generators (RNGs) play a crucial rule in many applications ranging from computational chemistry, biophysics and nuclear medicine[1] to quantum cryptography[2] and implementation of various computing applications and cryptographic systems utilized in modern digital communications[3]. The RNGs are categorized into two types namely pseudorandom number generators and physical random number generators (PRNG). In first type, pseudorandom numbers are generated via deterministic algorithms that use a single random seed whereas random phenomena such as thermal noise in resistors, photon noise, and frequency jitter of oscillators have been used as physical sources of entropy for realization of PRNGs along with some other post and pre-processing techniques[3].

It is known that sequences of pseudorandom numbers generated from the same deterministic seed are identical which degrades the reliability and level of security in systems which mainly depend on pseudorandom numbers. This shows the need of employing PRNGs to achieve irreproducible and unpredictable truly RNG. However, PRNG have been limited to much slower rates than pseudorandom RNGs due to limitations of the rate and power of the mechanisms for extracting bits from underlying physical source of randomness[3]. The slow generation rates of PRNG compared with high data rates of modern digital world applications are considered fundamental weakness of this type of RNGs. So, the realization of fast PRNGs becomes an active area of research in recent years.

The optical chaos generated by semiconductor lasers subject to various types of delayed feedback has attracted considerable interest during the last two decades[4-6]. From applications point of view, it has some interesting features such as broadband spectrum and extreme sensitivity to initial conditions. Thus, the high bandwidth optical chaos can be

---

employed to solve the problem of slow output rates of PRNGs and render the realization of ultra-fast PRNG possible as presented in recent literatures. The development of random number generators based on sampling the output of chaotic laser source has progressed recently in terms of generation rate as various schemes for random number generation have been reported since the first demonstration was carried out in 2008 at generation rate at 1.7 Gb/s[7] comparable to no more than 10 Mb/s typical rates in physical noise generators[7].

Since then, by employing several post and pre-processing techniques such as high-resolution analog to digital converters, high-order differential method, and bandwidth-enhanced chaos generation, the speed of PRNG has increased rapidly in the last 7 years. For example, the generation rate of random bits has been increasing very rapidly from the generation rate of 12.5 Gb/s in 2009[8] to the generation rate at 75 Gb/s in 2010[9] by employing the least significant bits (LSBs) technique in both RNGs. It is reported that high-order differential method increases the generation rate to 300 Gb/s[10] whereas the bit-order reversal method along with 8 LSBs sampled at 50 GHz brings up 400 Gb/s rate[11]. Recently, a generation rate of 2.2 Terabit per second (Tb/s) is reached via high-order finite differences pseudo RNG[12] in 2014 while a 1.4 Tb/s physical RNG was reported in 2015 using 2 chaotic data lines sampled at 100 GS/s by 8 bits ADC[13].

The value of time delay (TD) of feedback used in optical chaos generator represents one of the primary secret keys of chaotic system. An eavesdropper who can extract the TD value is, at least in principle, readily capable to reconstruct the optical chaos generator system. On the other hand, obvious TD signature directly diminishes the statistical performance of the PRNG. Therefore, it is essential to consider chaotic systems of suppressed TD signature in implementation of PRNGs.

The main objectives of this work is to propose and numerically demonstrate a novel multi-bit extraction scheme for generating ultra-fast physical random sequence of bits by utilizing very recent optical chaotic VCSEL[14] with suppressed TD signature in both intensity and phase of its two channel. The rest of the paper is organized as follows; the setup for the proposed PRNG is introduced in section 2, the mathematical model and numerical results are introduced in section 3, and finally section 4 contains the conclusion.

## 2. THR PROPOSED SETUP

The proposed technique uses an optical chaos generator of two output channels, based on single VCSEL diode, as depicted in Fig. 1. The two output channels (CH$_0$ and CH$_1$) are verified to be non-correlated and of a wide bandwidth[14]. By tuning the variable attenuator VA2, one guarantees the optical signal non-saturates the subsequent optical detection. The fast photodetector PD2 translates the optical intensity of the signal at CH$_0$ into an electrical signal which is then amplified using the amplifier Amp2.

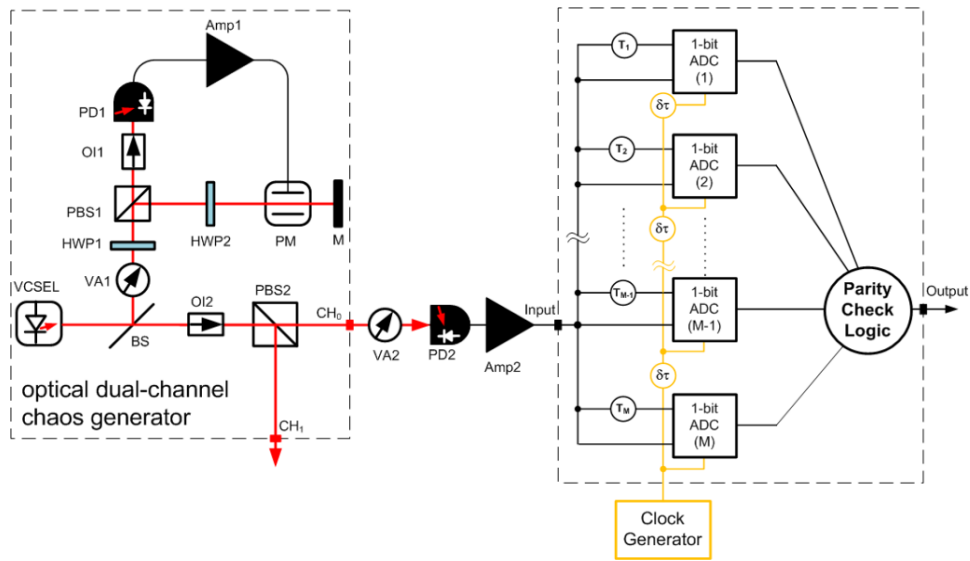

Figure 1. Schematic diagram of the ultra-fast physical random number generator using dual-channel optical chaos source. BS: beam splitter; VA: variable attenuator; HWP: half-wave plate; PBS: polarizing beam splitter; PD: photodetector; Amp: electro-optic gain; PM: phase modulator; M: mirror; OI: optical isolator; ADC: analog-to-digital converter.

The amplified chaotic intensity signal at time $t$, namely I($t$), is subjected to an array of fast 1-bit ADCs. At the $m$th 1-bit ADC, the signal I($t$) is compared with its delayed version $I(t - T_m)$ where $m = \{1, 2, ..., M\}$. The binary output of each 1-bit ADC is then sampled by a positive-edge-triggered D flip-flop that operates under the control of a clock signal at a rate $f_c = 1/\tau$ ($\tau$ is the clock period). The clock signal applied to the $m$th flip-flop is relatively delayed such that its rising edge precedes the edge of the clock of a subsequent $m$+1 flip-flop by a fixed period $\delta\tau = \tau/M$. The outputs of all D flip-flops are then combined by means of a parity-check logic, therefore its output is sensitive to the individual flipping of each 1-bit ADC. Here, the timing of the different delay units is a crucial point of design. The different values of $T_m$ are mutually incommensurable with

$$T_m > \tau, \text{ and } (T_i - T_j)_{i \neq j} > \tau$$

which is related to $B_w$; the bandwidth of chaotic signal of CH$_0$ as

$$\delta\tau = \frac{\tau}{M} \gg 1/B_w$$

It is shown that the generation rate of physically random bits is $M \times f_c$ which results in total key generation rate of $2 \times M \times f_c$, due to the dual-channel chaos source.

## 3. MATHEMATICAL MODEL AND NUMERICAL SIMULATIONS

The following rate equations describes the mathematical model of VCSEL based optical chaos generator[14] employed in this work.

$$\frac{dE_{x,y}}{dt} = k(1 + i\alpha)\{[N(t) - 1]E_{x,y}(t) \pm in(t)E_{y,x}\} \mp [g_a + g_p]E_{x,y}(t)$$
$$+ \sqrt{\beta_{sp}}\zeta_{x,y} + g_{1,2}\{\cos(2\theta_{p1})E_y(t - \tau_o)\sin(2\theta_{p1})E_x(t - \tau_o)\}[\cos^2(2\theta_{p2})\Psi_{x,y}$$
$$+ e^{i(t-\tau_e - 2\tau_{o1} - \tau_{o2})}\sin^2(2\theta_{p2})\Psi_{x,y}]e^{-i\omega_0 t}, \quad (1)$$

$$\frac{dN(t)}{dt} = g_N\{\mu - N(t)[1 + |E_x(t)|^2 + |E_y(t)|^2 + in(t)[E_x(t)\bar{E}_y(t) - \bar{E}_x(t)E_y(t)]\}, \quad (2)$$

$$\frac{dn(t)}{dt} = -g_s n(t) - g_N\{n(t)[|E_x(t)|^2 + |E_y(t)|^2 + iN(t)[E_y(t)\bar{E}_x(t) - \bar{E}_y(t)E_x(t)]\} \quad (3)$$

$$\phi(t) = |E_y(t)\sin(2\theta_{p1}) - E_x(t)\cos(2\theta_{p1})|^2, \Psi_x = \sin(2\theta_{p1}), \Psi_y = \cos(2\theta_{p1}) \quad (4)$$

where subscripts $x$ and $y$ stand for horizontal and vertical linear polarized (LP) modes, respectively, and the other parameters are described in Table 1.

We solve (1)-(4) using the following VCSEL parameters values[14]: k = 300 ns$^{-1}$, α = 4, $g_N$ = 1 ns$^{-1}$, $g_a$ = 0.5 ns$^{-1}$, $g_p$ = 30 ns$^{-1}$, $g_s$ = 50 ns$^{-1}$, $\beta_{sp}$ = 10$^{-6}$ ns$^{-1}$, μ = 4.5, and ω$_0$= 2.2176×10$^{15}$ rad/s. For simplicity, we take $g_1 = g_2 = g$, $\theta_{pi}$ = 22.5º and $\sqrt{\beta_{sp}}\zeta_x = \sqrt{\beta_{sp}}\zeta_y$ =0.

Theoretical study is employed to emphasize that firstly the TD signature is suppressed at chosen values of $T_m$s that are used in the proposed setup. The concealment of TD signature implies that the proposed PRNG is reliable and immune to any trial to extract information about its parameters. This goal can be verified using the well-known autocorrelation function (ACF) technique which has the advantage of being computationally efficient, robust, and immune to noise[15]. As the tendency of a given time series waveform to match its time-shifted version is quantified by ACF, the locations of peaks in ACF curve identify the presence of TD signature in chaotic output waveform.

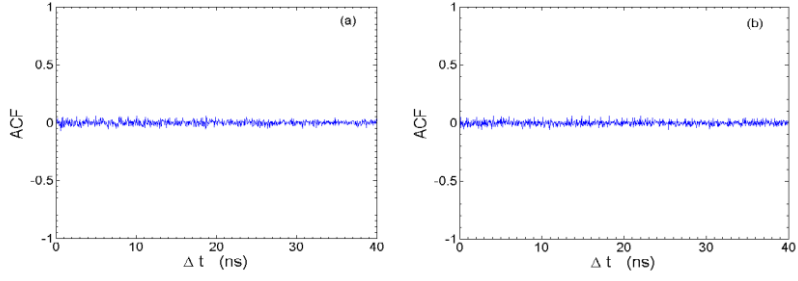

Figure 2. The ACF of intensity chaos of (a) x and (b) y linear polarized modes of system (1)-(4).

Numerical simulations are carried out using suitable values of parameters[14], i.e. $g$ = 15 ns$^{-1}$, $\tau_o$ = 6 ns, $\tau_e$ = 23.25 ns, and display that there is no particular values where significant peaks can be observed. In other words, there are no avoidable values of time delays employed in setup so that the chaotic time series is appropriate for generation of random bits utilizing both $x$ and $y$ output channels.

The values of time delays for chaotic signal $T_m$ are chosen as $T_m = \sqrt{3.2m - 1}$, $m$ =1, 2, …, $M$. Figure 2 illustrates the concealment of TD signature in chaotic output of both $x$ and $y$ channels which we hope for improving randomness quality of generated random sequences of bits.

Table 1. Description of the parameters of mathematical model (1)-(4)[14]

| Parameter | Description |
| --- | --- |
| E | Slowly varying complex amplitude of the electric field. |
| k | Cavity decay rate. |
| α | Linewidth enhancement factor. |
| $g_N$ | Decay rate of total carrier population. |
| N | Total carrier inversion between the conduction and valence bands. |
| n | Difference between carrier inversions of the spin-up and spin-down radiation channels. |
| $g_1$ and $g_2$ | Feedback strengths of the linear polarized modes. |
| $\tau_{01}$ | Delay period between VCSEL and PBS. |
| $\tau_{02}$ | optical delay period between PBS and M. |
| $\tau_0 = 2(\tau_{01} + \tau_{02})$ | Optical roundtrip time. |
| $\tau_e$ | Electronic time delay of the electro-optic feedback. |
| $g_a$ and $g_p$ | Linear anisotropies representing dichroism and birefringence, respectively. |
| $g_s$ | Spin-flip rate. |
| $\beta_{sp}$ | Spontaneous emission factor. |
| $\zeta_x$ and $\zeta_y$ | Gaussian white noises of zero mean value and unit variance. |
| $\mu$ | Normalized injection current with $\mu = 1$ at threshold. |
| $\omega_0$ | Center frequency of the solitary VCSEL. |

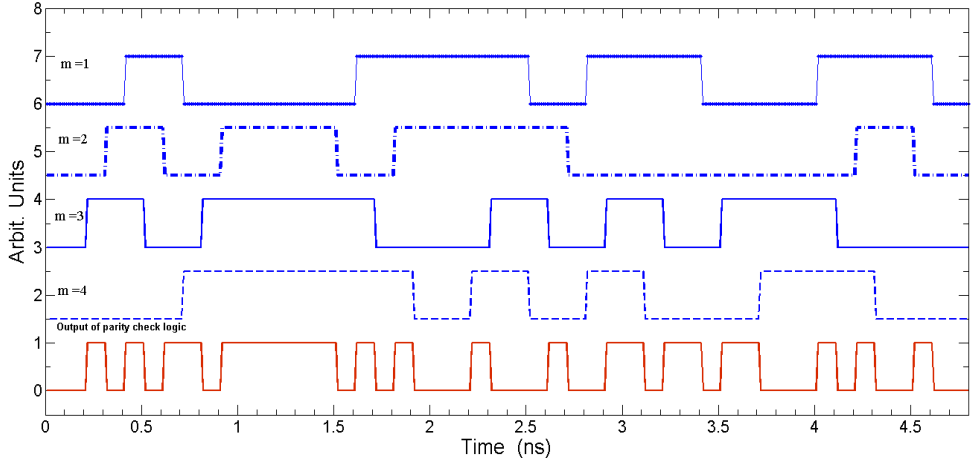

Figure 3. Snapshot of the output of four flip flops, in the first four rows, and its combined XOR result on the last row.

An example is given in Fig. 3 where the output of each flip flop at $m =1,2,3,4$ are shown in the first four rows and the generated random bits, obtained by logical XOR operation, are illustrated on the last row. The randomness quality of the generated bits is investigated using 15 standard statistical tests in NIST SP 800-22[15] at different values of $M$ in order to evaluate the maximum random bits generation rate of the proposed scheme. The NIST tests are conducted on 1000 1-Mbit sequences for a significance level of 0.01.

The results show that all tests are successfully passed from $M = 1$ to $M = 39$ at sampling rates extend from 10 up to 34.5 GHz, with the proportion of sequences satisfying p-value > 0.01 for 1000 samples of 1 Mbit data are in the range of $0.99 \pm 0.0094392$ as required[15]. This indicates that the maximum generation rate of random bits is 2.691 Tb/sec using one chaotic source of single VCSEL and without employing any pre-processing techniques. Increasing $M$ further degrades the performance of output random bits such that one or more of the statistical tests fails. The following table shows an example of tests results obtained at $M = 39$.

Table 2. Results of NIST statistical tests for generated physical random bits. For tests with multiple P-values, the worst case is shown.

| Number | Test | P-values |
|---|---|---|
| 1 | Monobit Frequency | 0.546642 |
| 2 | Block Frequency | 0.180407 |
| 3 | Runs | 0.843288 |
| 4 | Longest Runs Ones | 0.903194 |
| 5 | Binary Matrix Rank | 0.0246817 |
| 6 | Spectral | 0.247199 |
| 7 | Non-Overlapping Template Matching | 0.377646 |
| 8 | Overlapping Template Matching | 0.86798 |
| 9 | Universal Statistic | 0.263419 |
| 10 | Linear Complexity | 0.207562 |
| 11 | Serial | 0.256214 |
| 12 | Approximate Entropy | 0.272103 |
| 13 | Cumulative Sums | 0.772005 |
| 14 | Random Excursions | 0.416451 |
| 15 | Random Excursions Variant | 0.465517 |

## CONCLUSION

In conclusion, theoretical investigations illustrate that sequences of physical random bits generated by the proposed scheme have passed standard tests of randomness at ultra-fast rates up to 2.691 Tb/sec using dual channel chaotic laser source. To the best of our knowledge, the attained generation rate that we obtained is faster than any previously reported PRNG. The physical limitations of the optical chaos source, represented in bandwidth of chaos source, affect the rate of the generated physical random bits and make a maximum allowable value of $M$ be 39. So, further study employing broadband enhanced bandwidth optical chaos source can be investigated in future work to examine the possibility of increasing generation rate of physical random numbers.